\documentclass[conference]{IEEEtran}
\IEEEoverridecommandlockouts

\usepackage{cite}
\usepackage{amsmath,amssymb,amsfonts}
\usepackage{graphicx}
\usepackage{textcomp}
\usepackage{xcolor}
\usepackage{enumitem}
\usepackage{flushend}
\usepackage{booktabs}
\usepackage{multirow}
\usepackage{tabularx}
\usepackage{diagbox}
\usepackage{colortbl}
\usepackage{soul}
\usepackage{wrapfig}
\usepackage[caption=false]{subfig}
\usepackage[font=footnotesize,labelfont=bf]{caption}
\usepackage{dblfloatfix}
\usepackage{algorithm}
\usepackage{algcompatible}

\def\BibTeX{{\rm B\kern-.05em{\sc i\kern-.025em b}\kern-.08em
    T\kern-.1667em\lower.7ex\hbox{E}\kern-.125emX}}

\usepackage{comment}
\usepackage[caption=false]{subfig}
\usepackage[font=footnotesize,labelfont=bf]{caption}
\usepackage{tablefootnote}
\usepackage{makecell}
\usepackage{pifont}

\usepackage{flushend}
\usepackage{booktabs}
\usepackage{cite}
\usepackage{amsmath,amssymb,amsfonts}
\usepackage{algorithm}
\usepackage{graphicx}
\usepackage{textcomp}
\usepackage[colorlinks=true, urlcolor=black, linkcolor=black, citecolor=black, pdfborder={0 0 0}]{hyperref}

\definecolor{lightgreen}{HTML}{DFF0D8}
\definecolor{lightred}{HTML}{F2DEDE}
\definecolor{lightyellow}{HTML}{FCF8E3}

\usepackage{algpseudocode}

\begin{document}

\title{DecoRTL: A Run-time Decoding Framework for RTL Code Generation with LLMs}

\author{\IEEEauthorblockN{Mohammad Akyash, Kimia Azar, Hadi Kamali}
\IEEEauthorblockA{\textit{Department of Electrical and Computer Engineering (ECE), University of Central Florida, Orlando, FL 32816, USA} \\
\{mohammad.akyash, azar, kamali\}@ucf.edu}
}

\maketitle

\begin{abstract}

As one of their many applications, large language models (LLMs) have recently shown promise in automating register transfer level (RTL) code generation. However, conventional LLM decoding strategies, originally designed for natural language, often fail to meet the structural and semantic demands of RTL, leading to hallucinated, repetitive, or invalid code outputs. In this paper, we first investigate the root causes of these decoding failures through an empirical analysis of token-level entropy during RTL generation. Our findings reveal that LLMs exhibit low confidence in regions of structural ambiguity or semantic complexity, showing that standard decoding strategies fail to differentiate between regions requiring determinism (syntax-critical regions) and those that benefit from creative exploratory variability (design-critical regions). Then, to overcome this, we introduce DecoRTL, a novel run-time decoding strategy, that is both syntax-aware and contrastive for RTL code generation. DecoRTL integrates two complementary components: (i) self-consistency sampling, which generates multiple candidates and re-ranks them based on token-level agreement to promote correctness while maintaining diversity; and (ii) syntax-aware temperature adaptation, which classifies tokens by their syntactical and functional roles and adjusts the sampling temperature accordingly, enforcing low temperature for syntax-critical tokens and higher temperature for exploratory ones. Our approach operates entirely at inference time without requiring any additional model fine-tuning. Through evaluations on multiple open-source LLMs using the VerilogEval benchmark, we demonstrate significant improvements in syntactic validity, functional correctness, and output diversity, while the execution overhead (performance overhead) is imperceptible\footnote{Code is available at \NoHyper\cite{decortl_github}}.


\end{abstract}


\section{Introduction}

As hardware design becomes increasingly complex, machine learning (ML) offers a tempting path forward, enabling faster register transfer level (RTL) code generation, reducing manual design effort, and opening the field to a broader community of designers \cite{huang2021machine}. With  the rapid progress of large language models (LLMs), there is growing interest in leveraging them for hardware code synthesis from abstract design specifications \cite{wu2024chateda, chang2023chipgpt}. However, these models still face significant challenges in producing syntactically sound and semantically coherent RTL, particularly for structurally constrained and functionally complicated design scenarios. 

\begin{table}[b]
\centering
\footnotesize
\setlength{\tabcolsep}{3pt}
\caption{Comparison of RTL Code Generation Strategies with LLMs.}
\label{tab:methods_comparison}
\begin{tabular}{@{} l *{21}c @{}}
\toprule
\multirow{2}{*}{\textbf{Feature}} &  Prompt & Fine & \textbf{Decoding} \\
& Engineering & Tuning & \textbf{(Ours)} \\ 
\cmidrule(r){1-1} \cmidrule(r){2-2} \cmidrule(r){3-3} \cmidrule(r){4-4} 

Training Required & \cellcolor{lightgreen}No & \cellcolor{lightred}Yes & \cellcolor{lightgreen}No \\
\cmidrule(r){1-1} \cmidrule(r){2-2} \cmidrule(r){3-3} \cmidrule(r){4-4} 

Dataset Requirement & \cellcolor{lightgreen}Low & \cellcolor{lightred}High & \cellcolor{lightgreen}None \\
\cmidrule(r){1-1} \cmidrule(r){2-2} \cmidrule(r){3-3} \cmidrule(r){4-4} 

Compute Cost & \cellcolor{lightgreen}Low & \cellcolor{lightred}High & \cellcolor{lightgreen}Low \\
\cmidrule(r){1-1} \cmidrule(r){2-2} \cmidrule(r){3-3} \cmidrule(r){4-4} 

HW Engineer Knowledge Required & \cellcolor{lightyellow}Medium & \cellcolor{lightred}High & \cellcolor{lightgreen}Low \\
\cmidrule(r){1-1} \cmidrule(r){2-2} \cmidrule(r){3-3} \cmidrule(r){4-4} 

Adaptability to New Tasks & \cellcolor{lightyellow}Medium & \cellcolor{lightred}Low & \cellcolor{lightgreen}High \\
\cmidrule(r){1-1} \cmidrule(r){2-2} \cmidrule(r){3-3} \cmidrule(r){4-4} 

Effectiveness on Structural Constraints & \cellcolor{lightred}Low & \cellcolor{lightyellow}Medium & \cellcolor{lightgreen}High \\
\cmidrule(r){1-1} \cmidrule(r){2-2} \cmidrule(r){3-3} \cmidrule(r){4-4} 

Semantic Consistency & \cellcolor{lightred}Low & \cellcolor{lightyellow}Medium & \cellcolor{lightgreen}High \\
\cmidrule(r){1-1} \cmidrule(r){2-2} \cmidrule(r){3-3} \cmidrule(r){4-4} 

Output Diversity & \cellcolor{lightred}Low & \cellcolor{lightyellow}Medium & \cellcolor{lightgreen}High \\
\cmidrule(r){1-1} \cmidrule(r){2-2} \cmidrule(r){3-3} \cmidrule(r){4-4} 

Risk of Hallucination & \cellcolor{lightred}High & \cellcolor{lightyellow}Medium & \cellcolor{lightgreen}Low \\
\cmidrule(r){1-1} \cmidrule(r){2-2} \cmidrule(r){3-3} \cmidrule(r){4-4} 

Implementation Complexity & \cellcolor{lightgreen}Low & \cellcolor{lightred}High & \cellcolor{lightyellow}Medium \\
\cmidrule(r){1-1} \cmidrule(r){2-2} \cmidrule(r){3-3} \cmidrule(r){4-4} 

Reusability Across Models & \cellcolor{lightgreen}High & \cellcolor{lightred}Low & \cellcolor{lightgreen}High \\

\bottomrule
\end{tabular}
\end{table}

Current approaches to RTL code generation with LLM fall into two main categories \cite{akyash2024evolutionary}: (i) prompt engineering and (ii) fine-tuning. Prompt-based methods guide pre-trained models toward more accurate RTL generation through tailored instructions, task-specific context, or exemplars \cite{chang2023chipgpt, wu2024chateda}. In contrast, fine-tuning approaches focus on adapting LLMs using curated RTL datasets, either collected from open-source repositories \cite{akyash2025rtl++}, manually curated \cite{cui2024origen}, or synthesized from high-level specifications \cite{liu2024rtlcoder}, to expose models to a broader range of hardware design patterns. While both strategies are beneficial, they face critical limitations. Prompt engineering alone often lacks the robustness needed to consistently produce the needed (correct) RTL. On the other hand, fine-tuning demands large quantities of diverse, high-quality RTL data, which is difficult to collect and verify at scale. Despite these challenges, relatively little attention has been given to the decoding strategies used at inference time, which directly influences the quality, correctness, and diversity of LLMs' outputs, i.e., the generated RTL code. All existing LLM-based RTL generation methods rely on conventional decoding strategies, which is developed for natural language generation.

While standard decoding strategies have shown success in natural languages, they often fall short in code generation tasks \cite{zhu2024hot}, and RTL code is no exception. Unlike natural language, RTL code demands strict syntax/semantic correctness, structural precision with cycle-accurate concurrency, and an appropriate level of design diversity/creativity\cite{akyash2025simeval}. These limitations stems from two key issues: (i) The LLM with standard decoding often produces tokens with low confidence (i.e., high entropy), particularly in uncommon or complex regions of code, leading to unstable or incorrect outputs; (ii) In standard decoding with fixed temperature, the model fails to account for the varying needs of different token types in the code (syntax-critical tokens such as delimiters and keywords require high determinism to ensure validity, while semantically rich or design-critical tokens benefit from higher variability to support meaningful exploration of the design space).

\begin{figure}[t]
\centering
\includegraphics[width=\linewidth]{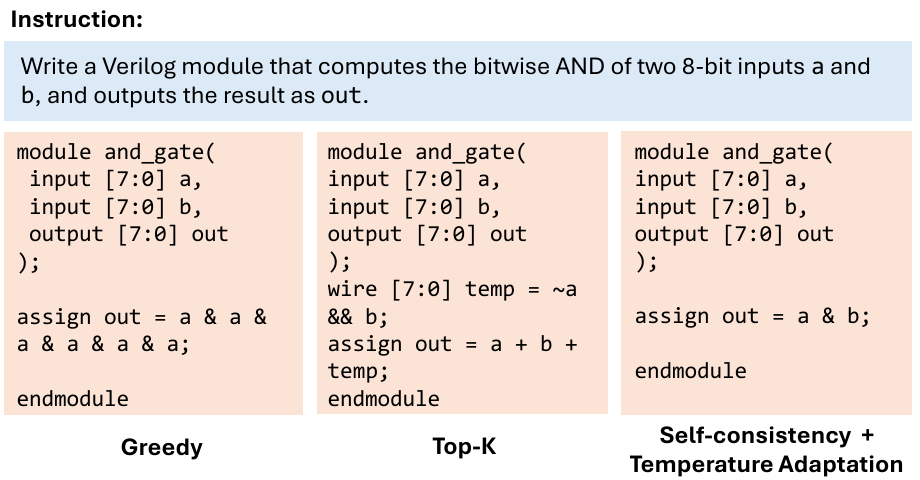}
\caption{RTL codes generated using different decoding strategies. Greedy and sampling-based methods often produce repetitive/invalid code, while our approach (Contrastive + TA decoding) generates more diverse and syntactically correct RTL \textbf{(Examples are abstracted due to space constrained)}\protect\footnotemark.}
\label{fig:methods_comparison}
\end{figure}

To address these limitations, the proposed framework, DecoRTL, takes a \textit{first-of-its-kind} step in shifting the focus of LLM-based RTL code generation from prompt engineering and fine-tuning to the unexplored space of decoding strategies, operating entirely at inference time. As summarized in Table~\ref{tab:methods_comparison}, and as shown in Fig. \ref{fig:methods_comparison} in an illustrative example, decoding can offer a unique set of advantages, such as zero training cost, high adaptability, and improved control over structural and semantic correctness, yet has received little attention in prior LLM-based RTL generation work. DecoRTL leverages contrastive plus temperature-adaptive (C\&TA) decoding, providing a lightweight and model-agnostic alternative that enhances output quality without modifying the underlying LLM. Specifically, our contributions are as follows:

\noindent \textbf{\ul{\emph{(1) Empirical Analysis of Uncertainty and Context:}}} We analyze the softmax distributions over logits during RTL decoding and observe that tokens appearing in ambiguous contexts often exhibit higher entropy, indicating model uncertainty and aligning with regions prone to structural or semantic errors. To further investigate, we categorize RTL tokens into high-impact and structural classes and analyze their context across open-source Verilog code. Our analysis shows that the class of a token is often predictable from its preceding token, revealing consistent local syntactic patterns in RTL.

\noindent \textbf{\ul{\emph{(2) Self-Consistency via Contrastive Top-K Re-ranking:}}} We propose an inference-time decoding method that improves self-consistency by selecting the top‑K candidate tokens at each decoding step and re‑ranking them via a contrastive mechanism, penalizing those overly similar to the mean embedding to improve diversity while preserving confidence and distinctiveness in token selection.

\noindent \textbf{\ul{\emph{(3) Syntax-Aware Temperature Adaptation:}}} Building on our entropy analysis, we introduce a dynamic temperature adaptation mechanism that predicts the class of the next token based on the most recently generated token. Using this prediction, we adjust the sampling temperature, applying lower temperatures for syntax-critical structural tokens to ensure determinism, and higher temperatures for high-impact tokens for diversity.

Through extensive experiments, we demonstrate the effectiveness of our domain-specific C\&TA decoding strategies. We evaluated our approach on multiple open-source LLMs using the VerilogEval benchmark \cite{liu2023verilogeval} and show that it consistently improves both the synthesizability and functional correctness of generated RTL code across all models. In particular, these improvements come with minimal computational overhead, as DecoRTL operates purely at the decoding level. Our approach achieves these gains without any fine-tuning, additional data gathering, or expensive retraining which denotes the practicality and generalizability of our method and make it a lightweight yet powerful solution for enhancing RTL code generation with existing models.

\section{Related Works}

\subsection{LLM for Code Generation}

LLMs have achieved remarkable success in code generation tasks across a wide range of high-level (software) programming languages. Models,  such as Codex \cite{chen2021evaluating}, CodeGen \cite{nijkamp2022codegen}, and AlphaCode \cite{li2022competition}, leverage massive pretraining on diverse code corpora and have demonstrated the ability to translate natural language instructions into syntactically correct and functionally coherent source code. 

\footnotetext{Greedy decoding is deterministic but often yields short, repetitive, and structurally flawed RTL code. Beam search \cite{Vijayakumar2018beam} improves coverage but tends to generate generic and redundant designs. Sampling methods like top-k \cite{fan2018hierarchical} add diversity through randomness, but in RTL generation, they can introduce syntax errors, semantic inconsistencies, and hallucinated hardware logic.}

Building on these capabilities, LLMs have recently been to a variety of hardware design and verification related tasks, including hardware debugging \cite{huang2024towards} design optimization \cite{yao2024rtlrewriter} (e.g., pipelining or parallelization), and detection of hardware-oriented security vulnerabilities \cite{akyash2024selfhwdebug, mashnoor2025llmift}. In parallel, several efforts have explored the use of LLMs for RTL code generation \cite{chang2023chipgpt, wu2024chateda, zhang2024mg, liu2024craftrtl, zhao2024codev}. Early approaches, such as ChatEDA \cite{wu2024chateda} and ChipGPT \cite{chang2023chipgpt}, employed prompt engineering techniques to guide general-purpose models (such as GPT-3.5 or GPT-4) by embedding design specifications, toolchain feedback, and format constraints directly into the prompts. These prompt-based strategies aim to elicit accurate Verilog outputs and capture design intent or tool responses, but often require manual intervention, prompt iteration, or post-processing to ensure valid and synthesizable outputs.

Beyond prompt engineering, more recent efforts have turned to fine-tuning open-source language models on RTL-specific datasets. VeriGen \cite{thakur2024verigen}, which utilized Verilog data from public repositories for supervised training, and RTLCoder \cite{liu2024rtlcoder}, which addressed dataset limitations by synthesizing instruction-code pairs uaing GPT3.5 to enhance diversity. Advancing this line of research, OriGen \cite{cui2024origen} introduced mechanisms like code augmentation and self-reflection to iteratively refine model outputs, while BetterV \cite{pei2024betterv} focused on design optimization by aligning generation objectives with Power, Performance, and Area (PPA) metrics. Additionally, CraftRTL \cite{liu2024craftrtl} enriched model understanding by integrating auxiliary design artifacts, e.g., state diagrams and waveforms into the training process.

While these methods have significantly advanced LLM-based RTL code generation, they share a common underlying limitation: they inherit natural language-oriented decoding strategies, which are not suited to the strict syntactic, structural, and semantic constraints of RTL code. Specifically, they may produce hallucinated logic, incomplete modules, or structurally invalid outputs, especially when the model encounters complex or uncommon design scenarios. Moreover, even in fine-tuned models, decoding is typically performed using greedy search, beam search, or top-$k$ sampling, which is not optimized for the unique demands of RTL. To date, no prior study has explored decoding-time adaptations as a means of improving RTL generation with LLMs. This is a critical gap in the field: while prompt engineering and fine-tuning rely on extensive human effort or large-scale dataset curation, decoding-time strategies offer a lightweight, generalizable, and model-agnostic alternative capable of significantly enhancing code quality without additional data collection or training.

\subsection{Decoding Strategies in LLMs}

LLMs generate text through an autoregressive decoding process, producing one token at a time based on the probability distribution conditioned on the previously generated context \cite{brown2020language}. While the model computes probabilities over the vocabulary, the actual output depends on the decoding strategy, which governs how tokens are selected. These strategies are broadly divided into the following categories: 

\noindent \textbf{\ul{\emph{(1) Deterministic Decoding:}}} Deterministic methods prioritize coherence and confidence by always selecting the most probable tokens. For instance, greedy decoding chooses the highest-probability token at each step, making it simple and fast but often resulting in repetitive or generic outputs \cite{song2024good}. Beam search \cite{Vijayakumar2018beam} improves upon this by maintaining multiple candidate sequences to optimize the overall sequence-level likelihood. However, it still tends to favor low-diversity completions due to its emphasis on probability maximization. 

\noindent \textbf{\ul{\emph{(2) Stochastic/Probabilistic Decoding:}}} Stochastic methods introduces controlled randomness to increase diversity and reduce repetition. Temperature sampling \cite{renze2024effect} adjusts the sharpness of the probability distribution: lower temperatures make the model more confident (favoring top tokens), while higher values flatten the distribution to allow more exploration. Top-$k$ sampling \cite{fan2018hierarchical} restricts sampling to the $k$ most likely tokens, while top-$p$ (nucleus) sampling \cite{holtzman2020curious} dynamically selects from the smallest set of tokens whose cumulative probability exceeds a threshold $p$, enabling adaptive diversity.

\noindent \textbf{\ul{\emph{(2) Constrastive Decoding:}}} Contrastive methods \cite{chuang2023dola} aims to improve generation quality by penalizing undesirable patterns rather than relying solely on probabilities. SimCTG \cite{su2022contrastive} implements this by discouraging tokens that are semantically similar to recent context embeddings, thereby reducing degeneration in natural language generation. Another variant \cite{li2023contrastive} uses two models, an “expert” and an “amateur”, to prefer tokens endorsed by the expert but disfavored by the amateur, enhancing reasoning and reducing hallucination \cite{o2023contrastive}. 

\begin{figure}[t]
\centering
\includegraphics[width=\linewidth]{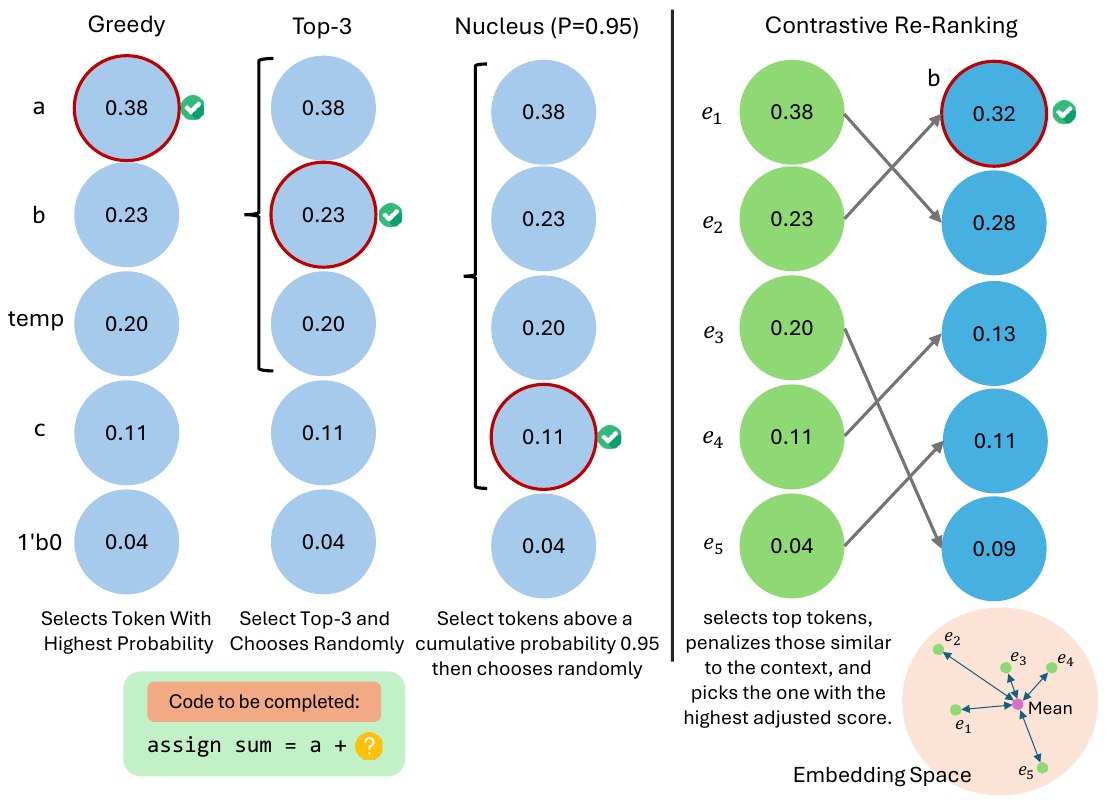}
\caption{The Overall Comparison of  Greedy, Top-$k$, Nucleus, and Our Contrastive Decoding Strategies for RTL Token Generation.}
\label{fig:different_decoding}
\end{figure}

As shown in Figure \ref{fig:different_decoding}, decoding strategies, whether deterministic, stochastic, or contrastive, vary in their implementation, provide trade-offs between  diversity, confidence, and syntactic control. While contrastive approaches have recently gained attention for their ability to improve generation diversity and consistency, they often rely on access to full hidden states, intermediate activations, or additional external models. These requirements limit their practicality and compatibility with structured, syntax-sensitive domains like RTL. In contrast, our approach introduces a lightweight, decoding-time contrastive mechanism that is fully model-agnostic and easily deployable across pre-trained LLMs. By applying token-level re-ranking within the top-$k$ candidates and penalizing options that are overly similar to the mean embedding, our method promotes output diversity without compromising syntactic fidelity or introducing architectural complexity.

\section{Uncertainty in RTL Code Generation w/ LLMs}
\label{sec:observation}

This section presents a set of key empirical observations that motivate our decoding strategy for RTL code generation with LLMs. Through comparative analysis with natural language (NL) generation and a corpus-level examination of RTL syntax patterns, we identify characteristic entropy dynamics and local context regularities that proves the need for a syntax-aware, temperature-adaptive decoding approach.

\begin{figure}[b]
\centering
\includegraphics[width=\linewidth]{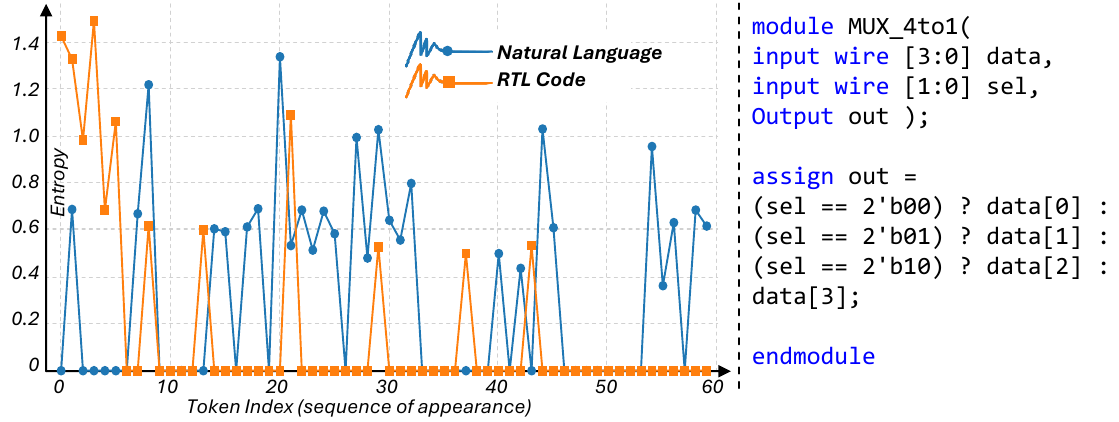}
\caption{Token-wise Entropy and their Comparison between Prompts for Natural Language vs. that of RTL Generation.}
\label{fig:entropy_code}
\end{figure}

\subsection{Entropy Patterns in Natural Language vs. RTL Generation}

To quantify the model’s confidence (or the level of uncertainty) during generation, we conducted a comparative experiment between the RTL code and the NL outputs. We prompted a quantized Qwen2.5 model with two tasks: (1) a free-form NL request, and (2) a prompt for generating a simple RTL (Verilog) code. At each task, and during the decoding step, we recorded the softmax distribution over the vocabulary, computed its entropy, and plotted the token‑wise entropy curves, as demonstrated in Fig.~\ref{fig:entropy_code}.

The analysis mentioned above reveals key differences: 

\noindent \textbf{\ul{\emph{(1) Higher Overall Uncertainty in NL:}}} As shown in Fig. \ref{fig:entropy_code}, the NL generation shows significantly higher mean entropy and greater variance (vs. RTL code generation), which reflects its inherently open-ended structure and the large number of plausible continuations at each step. 

\noindent \textbf{\ul{\emph{(2) Localized Entropy Spikes in RTL Generation:}}} While the RTL code shows overall lower entropy due to its rigid syntax, sharp entropy peaks occur at specific points, especially around control construsts (e.g., \texttt{if}, \texttt{always}, etc.) or block closures (e.g., \texttt{endif}, \texttt{endmodule}, etc.). These spikes reflect the potential ambiguity or uncommon contexts and correspond to regions with errors (e.g., missing or misplaced tokens). 

\noindent \textbf{\ul{\emph{(3) Entropy Decay Across the RTL Sequence:}}} At the beginning of the RTL code generation, e.g., modules headers, port declarations, signaling names, etc., entorpy is relatively high as the model explores naming and configuration options. As the code structure solidifies, entropy gradually decreases, showing growing model confidence. 

Such observations show the fragility of RTL code generation, where small errors in high-entropy regions (e.g., missing syntactical requirements) can make a module non-synthesizable or functionally incorrect. As such, identifying and mitigating these uncertainty peaks is critical to improving generation quality for the RTL code. 

\subsection{Structural Regularities in Token Contexts}

To further understand the structured nature of RTL code, we performed a corpus-level analysis on approximately 200K open-source Verilog modules. Our goal was to evaluate whether the local context preceding a token can predict its functional class (serves as an insight that informs our syntax-aware temperature adaptation mechanism. We focused on two classes of tokens in the RTL code (see Table \ref{tab:token_classes}): 

\begin{itemize}[leftmargin=*]
    \item \textbf{Structural Tokens:} (e.g., \texttt{endmodule}, \texttt{begin}, punctuations) that define code boundaries and syntax.

    \item \textbf{High-impact Tokens:} (e.g., operators, logic keywords) that carry significant semantic weight in hardware behavior.
\end{itemize}

\begin{figure}[t]
\centering
\includegraphics[width=\linewidth]{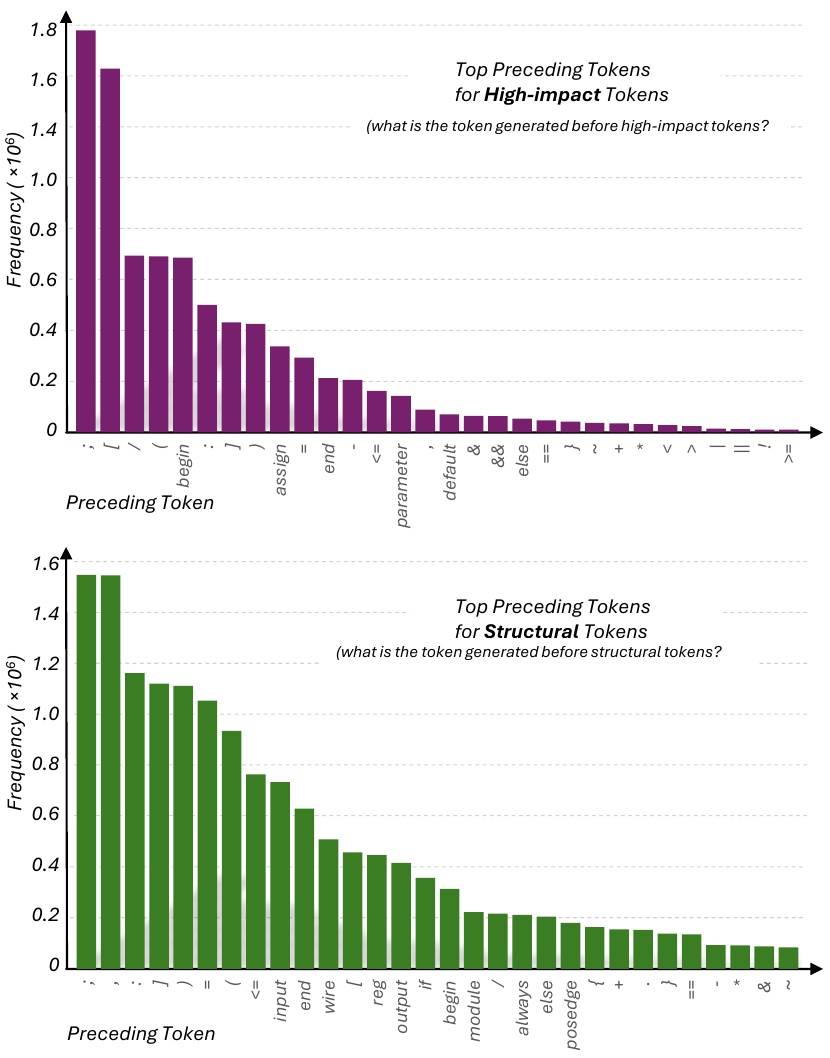}
\caption{Distribution of Preceding Tokens for Structural and High-Impact Categories \textbf{(For code generation purposes, only syntactic tokens—such as keywords, operators, and punctuation—are considered; variable names and identifiers are excluded from parsing.)}.}
\label{fig:token_contexts}
\end{figure}

For each token in these categories (listed in Table~\ref{tab:token_classes}), we computed the most frequent preceding tokens to assess context predictability. As shown in Fig.~\ref{fig:token_contexts}:

\begin{itemize}[leftmargin=*]
    \item \textit{Structural tokens are typically preceded by control keywords or delimiters, indicating deterministic, syntax-sensitive regions where precise decoding is essential.}

    \item \textit{High-impact tokens more often follow expressions, operands, or operators, reflecting semantically rich regions where exploration and diversity may yield better designs.}
\end{itemize}

These patterns confirm that the context of the most recently token provides a strong signal for adjusting decoding behavior. Structural tokens benefit from low-temperature (deterministic) decoding to maintain validity, while high-impact tokens can tolerate or even benefit from increased sampling temperature to support design variability. Together, these observations provide the foundation for our decoding-time framework, which dynamically adapts sampling temperature and contrastive selection based on local context and token uncertainty.

\begin{table}[t]
\centering
\caption{Token classes used for syntax-aware temperature adaptation.}
\label{tab:token_classes}
\fontsize{8pt}{9pt}\selectfont
\setlength\tabcolsep{3pt}
\begin{tabular}{@{}p{0.14\textwidth}p{0.30\textwidth}@{}}
\toprule
\textbf{Token Class} & \textbf{Tokens} \\
\midrule
High-impact &
\texttt{+}, \texttt{-}, \texttt{*}, \texttt{/}, \texttt{\&}, \texttt{|}, \texttt{\^}, \texttt{\~}, \texttt{!}, \texttt{=}, \texttt{==}, \texttt{!=}, \texttt{<}, \texttt{<=}, \texttt{>}, \texttt{>=}, \texttt{?}, \texttt{:}, \texttt{<=}, \texttt{=>}, \texttt{\&\&}, \texttt{||} \\
\addlinespace[1pt]
Structural &
\texttt{module}, \texttt{endmodule}, \texttt{input}, \texttt{output}, \texttt{inout}, \texttt{wire}, \texttt{reg}, \texttt{logic}, \texttt{parameter}, \texttt{assign}, \texttt{always}, \texttt{begin}, \texttt{end}, \texttt{if}, \texttt{else}, \texttt{case}, \texttt{default}, \texttt{for}, \texttt{while}, \texttt{;}, \texttt{,}, \texttt{.}, \texttt{[}, \texttt{]}, \texttt{(}, \texttt{)}, \texttt{\{}, \texttt{\}}, \texttt{posedge}, \texttt{negedge} \\
\bottomrule
\end{tabular}
\end{table}

\section{Proposed Method: DecoRTL}

\begin{figure*}[t]
\centering
\includegraphics[width=\textwidth]{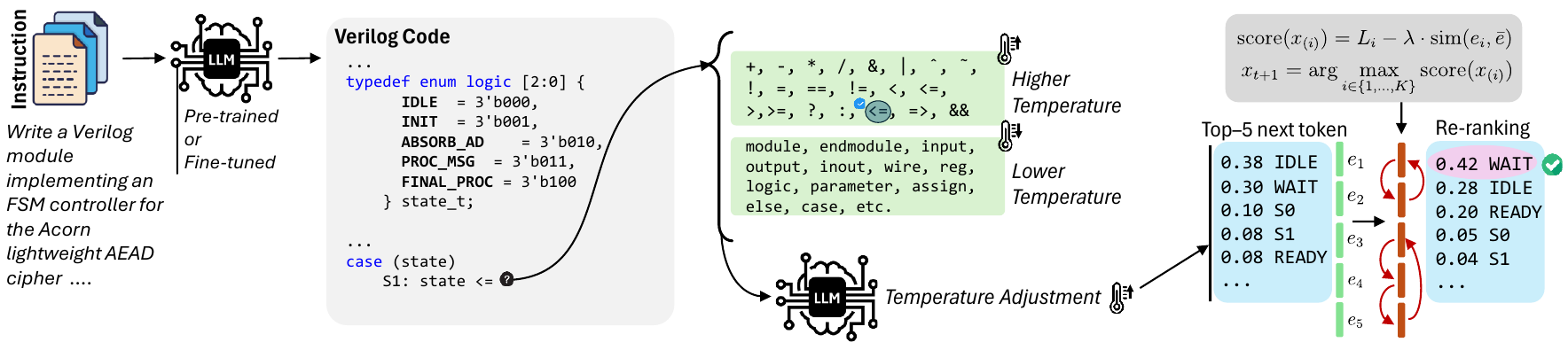}
\caption{Overview of DecoRTL Framework for RTL Generation. Given a design prompt (here the encryption controller FSM), the LLM generates RTL code (token by token) with two decoding-time enhancements: syntax-aware temperature adaptation adjusts sampling based on token type (the middle part), and contrastive re-ranking selects the next token by penalizing semantic redundancy among top candidates (the right part), improving RTL quality w/o retraining.}
\label{fig:overview_framework}
\end{figure*}

Fig. \ref{fig:overview_framework} depicts an overview of the DecoRTL framework, illustrated through an example in which an LLM is tasked with generating the controller logic of a lightweight authenticated encryption with associated data (AEAD) module. Our proposed domain-specific decoding strategy to improve RTL code generation consists of two parts: (i) \textbf{self-consistency re-ranking}, which promotes stability and output diversity by re-evaluating top-$k$ token candidates based on embedding similarity, and (ii) \textbf{syntax-aware temperature adaptation}, which dynamically modulates the sampling temperature according to the predicted class of the next token. As illustrated in Fig. \ref{fig:overview_framework}, these components operate entirely at inference time and are designed to enhance both the syntactic validity and functional correctness of generated RTL code. 

\subsection{Contrastive Self-Consistency Decoding}

In standard autoregressive decoding, a language model generates a sequence \(X = (x_1, x_2, \dots, x_t)\) one token at a time. At each time step \(t\), the model outputs a vector of logits \(\mathbf{z}^{(t)} \in \mathbb{R}^{|V|}\) over the vocabulary \(V\). After applying temperature scaling with parameter \(T\), the probability of token \(x\) given the generated context \(x_{1:t}\) is computed using the softmax function:
\[
p(x \mid x_{1:t}) = \frac{\exp(z_x^{(t)}/T)}{\sum_{y \in V}\exp(z_y^{(t)}/T)}
\]

While this formula prioritizes high-likelihood tokens, it may lead to repetitive or semantically clustered outputs, which is error-prone in structured domains like RTL code. To address this, we introduce a contrastive re-ranking mechanism that promotes self-consistency and diversity among high-probability candidates. Our method refines token selection by using embedding-level similarity among the top-$K$ candidates. First, we extract the top-$K$ tokens from the probability distribution:
\[
\{x_{(1)}, x_{(2)}, \dots, x_{(K)}\},
\]
with corresponding log-probabilities:
\[
L_i = \log p(x_{(i)} \mid x_{1:t}), \quad i = 1, \dots, K.
\]

Each candidate \(x_{(i)}\) is mapped to its embedding \(e_i\) via the model's embedding matrix \(E\) as $e_i = E(x_{(i)}),$ and normalized so that \(\|e_i\| = 1\). The mean embedding of the top-$K$ candidates is computed as:
\[
\bar{e} = \frac{1}{K} \sum_{i=1}^{K} e_i.
\]

This mean serves as a semantic reference point for evaluating diversity. Then we calculate the cosine similarity between each candidate embedding and the mean is given by:
\[
\text{sim}(e_i, \bar{e}) = e_i \cdot \bar{e},
\]
where a higher similarity indicates a candidate is closer to the average, and thus more redundant.

To promote diversity, we penalize candidates that are too close to this mean. The adjusted score is defined as:
\[
\text{score}(x_{(i)}) = L_i - \lambda \cdot \text{sim}(e_i, \bar{e}),
\]
where \(\lambda\) is a tunable hyperparameter controlling the trade-off between confidence and diversity. he next token is then selected as the candidate with the highest adjusted score:
\[
x_{t+1} = \arg\max_{i \in \{1, \dots, K\}} \text{score}(x_{(i)}).
\]

This re-ranking strategy ensures that the selected token is both probabilistically sound and semantically distinct from the average of the high-probability set. From a probabilistic perspective, this modification can be viewed as adjusting the original distribution with a diversity-aware term. The original probability for a top-$K$ candidate is:
\[
p(x_{(i)} \mid x_{1:t}) = \frac{\exp(L_i)}{Z},
\]
where \(Z\) is the partition function. After applying the contrastive penalty, the unnormalized modified probability is:
\[
p_{\text{modified}}(x_{(i)} \mid x_{1:t}) \propto p(x_{(i)} \mid x_{1:t}) \cdot \exp(-\lambda \cdot \text{sim}(e_i, \bar{e})).
\]

This re-weighting flattens overconfident, high-density regions in the distribution, where semantically redundant candidates dominate, and redistributes probability mass toward diverse, yet still plausible tokens. As a result, it reduces the dominance of clustered tokens and produces outputs that are not only high-quality and coherent but also less prone to repetition and syntactic collapse.

In LLMs, tokens with similar semantic and syntactic roles naturally cluster together in the embedding space. Thus, the top-K candidates often represent minor variations on a common theme. By penalizing candidates with high cosine similarity to the mean embedding, our method discourages the model from repeatedly choosing near-duplicate tokens. This helps to produce a sharper, more definitive distribution, lowering the overall entropy, and promotes diversity by encouraging the selection of tokens that are distinct yet still plausible. The resulting output is both more stable and less susceptible to errors, which is critical in synthesizable RTL code where even a small mistake may render a design non-functional.

\subsection{Syntax-aware Temperature Adaptation}

In standard decoding, the temperature parameter $T$ controls the entropy of the output distribution. Lower temperatures make the output more deterministic, while higher temperatures introduce more randomness and diversity. However, a fixed temperature throughout decoding fails to reflect the varying demands of different token types in RTL code where certain tokens require precise syntax, while others benefit from exploratory flexibility. In DecoRTL, we introduce a \textit{syntax-aware temperature adaptation} strategy that dynamically adjusts the sampling temperature at each decoding step based on the syntactic role of the \textit{next} token. Rather than predicting token type directly, we infer it from the most recently generated token by leveraging structural patterns common in RTL. This approach is motivated by our earlier analysis (refer to Section \ref{sec:observation}), which demonstrated that certain token categories are frequently preceded by consistent syntactic patterns.

Based on the observation, we define two token classes:

\noindent \textbf{\ul{\emph{(1) Structural Tokens:}}} Tokens that define the syntactic framework of Verilog code and must be generated with high determinism to maintain correctness. These include keywords that establish module structure (\texttt{module}, \texttt{endmodule}, \texttt{input}, \texttt{output}, \texttt{wire}, etc.), control flow constructs (\texttt{if}, \texttt{else}, \texttt{case}, \texttt{for}, \texttt{while}), and delimiters or punctuation symbols (\texttt{;}, \texttt{,}, \texttt{.}, \texttt{[}, \texttt{]}, \texttt{(}, \texttt{)}, \texttt{\{}, \texttt{\}}).

\noindent \textbf{\ul{\emph{(1) High-impact Tokens:}}} These tokens contribute directly to the functional and logical semantics of a design and benefit from moderate sampling diversity to allow expression of valid alternatives. This category includes arithmetic and bitwise operators (\texttt{+}, \texttt{-}, \texttt{*}, \texttt{/}, \texttt{\&}, \texttt{|}, \texttt{\^}, \texttt{\~}), logical and comparison operators (\texttt{==}, \texttt{!=}, \texttt{<}, \texttt{<=}, \texttt{>}, \texttt{>=}, \texttt{\&\&}, \texttt{||}), and conditional tokens (\texttt{?}, \texttt{:}, \texttt{=}, \texttt{<=}, \texttt{=>}).

All remaining tokens do not require explicit control, including filler tokens in comments, identifiers in unambiguous contexts, or tokens in low-impact sections where variation does not significantly affect syntax or semantics.

This adaptive decoding strategy dynamically modulates the model’s sampling temperature based on the \textit{anticipated syntactic role} of the next token. Given a token \( x_t \) generated at time step \( t \), we estimate the likely category of the upcoming token \( x_{t+1} \) as \( C(x_{t+1}) \), and adjust the temperature accordingly:
\[
T_{t+1} =
\begin{cases}
T_{\text{base}} - 0.1, & \text{if } C(x_{t+1}) \text{ is structural}, \\
T_{\text{base}} + 0.1, & \text{if } C(x_{t+1}) \text{ is high-impact}, \\
T_{\text{base}},       & \text{otherwise.}
\end{cases}
\]

This dynamic temperature is then applied to the softmax distribution when generating the token at time \( t+1 \):
\[
p(x \mid x_{1:t}; T_{t+1}) = \frac{\exp(z_x / T_{t+1})}{\sum_{y \in V} \exp(z_y / T_{t+1})}.
\]

Lower temperatures sharpen the distribution, yielding more deterministic outputs, while higher temperatures flatten it, promoting diversity. As a result, the decoder becomes more conservative in syntax-critical contexts and more exploratory in regions where semantic flexibility is beneficial.

\begin{algorithm}[t]
\footnotesize
\caption{Token Generation via Contrastive Re-Ranking and Syntax-Aware Temperature Adaptation}

\begin{algorithmic}[1]
\State \textbf{Input:} Instruction $I$, decoder $d$, initial temperature $\tau_0$, penalty $\lambda$;
\State Initialize sequence $T \leftarrow \emptyset$, temperature $\tau \leftarrow \tau_0$;
\While{not end-of-sequence}

    \STATEx \hrulefill
    
    \STATEx \textbf{Step 1: Top-$K$ Candidate Generation}
    \State \hspace{1em} Get logits from decoder $d$;
    \State \hspace{1em} Apply temperature $\tau$;
    \State \hspace{1em} Extract top-$K$ tokens $\{x_{(1)}, \dots, x_{(K)}\}$;
    \State \hspace{1em} Extract log-probabilities $\{L_1, \dots, L_K\}$;

    \STATEx \hrulefill

    \STATEx \textbf{Step 2: Contrastive Re-Ranking}
    \State \hspace{1em} Normalize embeddings: $e_i \leftarrow \frac{E(x_{(i)})}{\|E(x_{(i)})\|}$; 
    \State \hspace{1em} Compute mean: $\bar{e} \leftarrow \frac{1}{K} \sum e_i$;
    \State \hspace{1em} Score: $\text{score}_i \leftarrow L_i - \lambda \cdot (e_i \cdot \bar{e})$;
    \State \hspace{1em} Select: $x^* \leftarrow \arg\max_i \text{score}_i$; Append $x^*$ to $T$;

    \STATEx \hrulefill
    
    \STATEx \textbf{Step 3: Syntax-Aware Temperature Adaptation}
    \If{next token likely structural (based on $x^*$)}
        \State \hspace{1em} $\tau \leftarrow \tau_0 - 0.1$;
    \ElsIf{next token likely high-impact}
        \State \hspace{1em} $\tau \leftarrow \tau_0 + 0.1$;
    \Else
        \State \hspace{1em} $\tau \leftarrow \tau_0$;
    \EndIf
\EndWhile
\State \textbf{Output:} Generated token sequence $T$;
\end{algorithmic}
\label{alg:overview_alg}
\end{algorithm}

Algorithm \ref{alg:overview_alg} shows the detailed decoding procedure of the \textbf{DecoRTL} framework, which integrates both \textit{contrastive re-ranking} and \textit{syntax-aware temperature adaptation} to enhance RTL code generation. Starting with initial temperature $\tau_0$, and a contrastive penalty parameter $\lambda$, the temperature variable $\tau$ is dynamically adjusted throughout the decoding process based on the local context of the generated tokens. Each decoding iteration consists of three main steps:  (1) The decoder generates logits at the current temperature $\tau$, and the top-$K$ candidate tokens $\{x_{(1)}, \dots, x_{(K)}\}$ are selected along with their corresponding log-probabilities $\{L_1, \dots, L_K\}$; (2) Each token’s embedding is compared to the mean embedding of the top-$K$ candidates using cosine similarity, $\text{sim}(e_i, \bar{e})$, which quantifies semantic redundancy. Tokens more similar to the mean are penalized, and the token with the highest adjusted score is selected; (3) The temperature $\tau$ is updated based on the class of the most recently generated token. If the token is structural, the temperature is decreased ($\tau \leftarrow \tau_0 - 0.1$) to enforce deterministic sampling. If the token is high-impact, the temperature is increased ($\tau \leftarrow \tau_0 + 0.1$) to encourage exploration. For all other token types (neither structural nor high-impact), the temperature is reset to its base value ($\tau \leftarrow \tau_0$). This decoding procedure enables \textbf{DecoRTL} to dynamically balance syntactic precision and semantic diversity, improving quality without requiring fine-tuning or more training.

\section{Experimental Results}

To assess the effectiveness of our proposed decoding strategies for RTL generation, we conduct a series of quantitative experiments across multiple dimensions: generation stability based on the decoding strategy, functional correctness (and synthesizability) w.r.t. the decoding, and decoding efficiency. We leverage the VerilogEval benchmark (Human) \cite{liu2023verilogeval} to evaluate models on realistic RTL design prompts. A diverse set of LLMs, either pre-trained or fine-tuned for RTL generation, varying in parameter size and architectural design, is used to ensure broad applicability. This includes QwenCoder-2.5-14B \cite{hui2024qwen2}, CodeLlama-7B \cite{roziere2023code}, and CodeV \cite{zhao2024codev}, and the following demonstrates the consistent improvements enabled by our decoding framework over these LLMs.

\subsection{Entropy-Based Evaluation of Decoding Methods}

To evaluate our decoding method w.r.t. generation stability and model confidence, we conducted an experiment analyzing how different decoding strategies affect token-level entropy during Verilog code generation. Entropy reflects the uncertainty of the model's prediction, higher values show indecision, while lower entropy suggests confident and consistent output. 

We began by sampling 40 instruction-to-code prompts from the VerilogEval \cite{liu2023verilogeval} and used them to generate Verilog modules with two LLMs: QwenCoder-2.5 (14B) and CodeLlama (7B). Each prompt was decoded using three strategies: (1) top-$k$ sampling with $k=10$, (2) nucleus sampling with $p=0.9$, and (3) our contrastive decoding method with $k=5$ and a penalty coefficient $\lambda = 0.5$. For each method, we computed the entropy of the softmax distribution at each generation step and aggregated the mean and variance across all prompts.

As shown in Table~\ref{tab:entropy_comparison}, contrastive decoding consistently achieves a lower mean and variance in entropy compared to both top-$k$ and nucleus sampling. A lower mean entropy indicates that the model is making more confident token selections, while a reduced entropy variance suggests that the model's confidence remains stable throughout the sequence. By flattening entropy spikes and narrowing the distribution of uncertainty, our decoding method improves the consistency and reliability of generated Verilog code, making it more aligned with the deterministic nature of RTL codes.

\begin{table}[t]
\centering
\footnotesize
\caption{Mean and Variance of Token-level Entropy during RTL Code Generation across Decoding Strategies.}
\label{tab:entropy_comparison}
\begin{tabular}{lcccc}
\toprule
\textbf{Strategy} 
& \multicolumn{2}{c}{\textbf{QwenCoder-2.5-14B}} 
& \multicolumn{2}{c}{\textbf{CodeLlama-7B}} \\
\cmidrule(lr){2-3} \cmidrule(lr){4-5}
& \textbf{Mean} & \textbf{Variance} & \textbf{Mean} & \textbf{Variance} \\
\midrule
Top-$k$ (k=10)        & 0.106 & 0.065 & 0.219 & 0.139 \\
\cmidrule(lr){1-1} \cmidrule(lr){2-3} \cmidrule(lr){4-5}
Nucleus (p=0.9)       & 0.176 & 0.144 & 0.275 & 0.179 \\
\cmidrule(lr){1-1} \cmidrule(lr){2-3} \cmidrule(lr){4-5}
\textbf{Contrastive (Ours)}   & \textbf{0.071} & \textbf{0.061} & \textbf{0.134} & \textbf{0.097} \\
\bottomrule
\end{tabular}
\end{table}

\begin{table}[t]
\centering
\footnotesize
\caption{Functionality Correctness Comparison of Fixed vs. Adaptive Temperature Decoding (The Model is CodeLlama 7B).}
\label{tab:temp_functionality}
\scriptsize
\begin{tabular}{lc}
\toprule
\textbf{Decoding Strategy} & \textbf{Functional Pass Rate (\%)} \\
\midrule
Fixed Temperature (Temperature set to $T = 0.5$) & 18.5 \\
\cmidrule(lr){1-1} \cmidrule(lr){2-2}
Fixed Temperature (Temperature set to $T = 0.7$) & 18.5 \\
\cmidrule(lr){1-1} \cmidrule(lr){2-2}
Fixed Temperature (Temperature set to $T = 0.9$) & 19.2 \\
\cmidrule(lr){1-1} \cmidrule(lr){2-2}
\textbf{Adaptive Temperature (Ours)} & \textbf{25.6} \\
\bottomrule
\end{tabular}
\end{table}

\subsection{Comparison with Fixed Temperature Decoding}

To evaluate the effectiveness of our syntax-aware adaptive temperature mechanism, we compare it against fixed-temperature decoding strategies using the CodeLlama 7B model. Specifically, we test three static temperature settings: $T = 0.5$, $T = 0.7$, and $T = 0.9$, representing conservative, balanced, and exploratory decoding behaviors, respectively. For each temperature, we generate RTL code from a shared set of design instructions and evaluate the functionality of the output based on a pass/fail criterion.

\begin{table}[b]
\centering
\footnotesize
\setlength\tabcolsep{3pt}
\caption{Number of Hallucinated and Repetitive RTL Code Outputs before and after Applying Contrastive + TA Decoding.}
\label{tab:hallucination_repetition}
\begin{tabular}{lcccc}
\toprule
\multirow{2}{*}{\textbf{Model}} 
& \multicolumn{2}{c|}{\textbf{Baseline (top-$k$)}} 
& \multicolumn{2}{c}{\textbf{Contrastive + TA}} \\
\cmidrule(lr){2-3} \cmidrule(lr){4-5}
& Hallucinated & Repetitive & Hallucinated & Repetitive \\
\cmidrule(lr){1-1}\cmidrule(lr){2-3} \cmidrule(lr){4-5}
CodeLlama 7B       & 18 & 9  & 3   & None \\
\cmidrule(lr){1-1}\cmidrule(lr){2-3} \cmidrule(lr){4-5}
QwenCoder-2.5 14B  & 11 & 5  & None & None \\
\bottomrule
\end{tabular}
\label{tab:hallucinated_rep}
\end{table}

As shown in Table \ref{tab:temp_functionality}, in fixed-temperature decoding, the model applies a uniform degree of randomness throughout generation, regardless of token context. Lower temperatures (e.g., $T=0.5$) bias the model toward high-confidence predictions, yielding deterministic but often repetitive and rigid code. In contrast, higher temperatures (e.g., $T=0.9$) promote diversity and exploration, but increase the likelihood of producing structurally invalid or semantically inconsistent outputs. Note that from a modeling perspective, adaptive temperature control better aligns with the internal behavior of transformer-based LLMs. During generation, different layers and attention heads specialize in capturing hierarchical and positional dependencies, especially important in structured domains like code. A fixed-temperature setting fails to leverage this internal structure, applying uniform sampling across both high-confidence and uncertain regions. 

\subsection{Reduction of Hallucination and Repetition in RTL Code}

To evaluate the impact of our full decoding framework on RTL code quality, we analyze the outputs of both LLMs, CodeLlama 7B and QwenCoder-2.5 14B, before and after applying our combined contrastive + temperature-adaptive decoding method. As shown in Table~\ref{tab:hallucination_repetition}, the baseline top-$k$ decoding strategy exhibits a considerable number of both hallucinated\footnote{Hallucination in RTL code refers to the generation of syntactically valid but semantically incorrect or meaningless hardware constructs that do not correspond to the design intent, including illogical control flows.} and repetitive\footnote{Repetition refers to the unintended duplication of RTL code segments, such as repeated logic, redundant declarations, or loops of identical expressions.} outputs. This is while applying our proposed decoding framework reduces these incidents. This highlights the complementary strengths of contrastive decoding and temperature adaptation: the former penalizes semantically redundant token choices, encouraging meaningful variation, while the latter enforces deterministic behavior in structurally sensitive regions and allows controlled exploration in semantically rich segments.

\begin{table}[t]
\centering
\setlength\tabcolsep{2pt}
\caption{RTL Generation Synthesizability Success Rate across Models.}
\label{tab:method_comparison_synth}
\begin{tabular}{@{} l *{21}c @{}}
\toprule
\textbf{Model} & \multicolumn{3}{c}{\textbf{CodeLlama 7B}} & \multicolumn{3}{c}{\textbf{QwenCoder 14B}} & \multicolumn{3}{c}{\textbf{CodeV \cite{zhao2024codev}}} \\
 & \multicolumn{3}{c}{\textbf{(Pre-trained)}} & \multicolumn{3}{c}{\textbf{(Pre-trained)}} & \multicolumn{3}{c}{\textbf{(RTL Fine-tuned)}} \\

\cmidrule(lr){2-4} \cmidrule(lr){5-7} \cmidrule(lr){8-10}
& Syn & Syn & Syn & Syn & Syn & Syn & Syn & Syn & Syn \\
& @1$^+$ & @5 & @10 & @1 & @5 & @10 & @1 & @5 & @10 \\
\cmidrule(lr){1-1} \cmidrule(lr){2-4} \cmidrule(lr){5-7}  \cmidrule(lr){8-10} 
Base$^*$ & 35.2\% & 38.4\% & 41.6 \% & 64.1\% & 70.5\% & 76.2\% & 75.6\% & 82.6\% & 85.9\% \\
\cmidrule(lr){1-1} \cmidrule(lr){2-4} \cmidrule(lr){5-7}  \cmidrule(lr){8-10} 
TA$^*$ & 41.6\% & 44.2\% & 47.4\% & 75.0\% & 79.4\% & 83.9\% & 83.3\% & 87.1\% & 90.3\%\\
\cmidrule(lr){1-1} \cmidrule(lr){2-4} \cmidrule(lr){5-7}  \cmidrule(lr){8-10}
C$^*$ & 42.3\% & 46.1\% & 49.3\% & 74.3\% & 78.8\% & 84.6\% & 82.6\% & 87.8\% & 89.1\% \\
\cmidrule(lr){1-1} \cmidrule(lr){2-4} \cmidrule(lr){5-7}  \cmidrule(lr){8-10} 
C+TA$^*$  & 46.1\% & 49.3\% & 52.5\% & 80.4\% & 82.0\% & 85.9\% & 88.4\% & 91.6\% & 93.5\% \\
\bottomrule
\multicolumn{13}{l}{$^*$: Base: Baseline decoding using top-$k$, TA: Temperature Adaptive Only,} \\
\multicolumn{13}{l}{~~~C: Contrastive Only, and C+TA: Contrastive with Temperature Adaptive.} \\
\multicolumn{13}{l}{$^+$: Syn@i means synthesizability in ``i'' runs.} \\
\end{tabular}
\end{table}

\begin{table}[t]
\centering
\setlength\tabcolsep{2pt}
\caption{RTL Generation Functional Correctness Rate across Models. }
\label{tab:method_comparison_func}
\begin{tabular}{@{} l *{21}c @{}}
\toprule
\textbf{Model} & \multicolumn{3}{c}{\textbf{CodeLlama 7B}} & \multicolumn{3}{c}{\textbf{QwenCoder 14B}} & \multicolumn{3}{c}{\textbf{CodeV \cite{zhao2024codev}}} \\
 & \multicolumn{3}{c}{\textbf{(Pre-trained)}} & \multicolumn{3}{c}{\textbf{(Pre-trained)}} & \multicolumn{3}{c}{\textbf{(Fine-tuned for RTL)}} \\

\cmidrule(lr){2-4} \cmidrule(lr){5-7} \cmidrule(lr){8-10}
& Pass & Pass & Pass & Pass & Pass & Pass & Pass & Pass & Pass \\
& @1$^+$ & @5 & @10 & @1 & @5 & @10 & @1 & @5 & @10 \\
\cmidrule(lr){1-1} \cmidrule(lr){2-4} \cmidrule(lr){5-7} \cmidrule(lr){8-10}
Base$^*$ & 18.2\% & 22.7\% & 24.3\% & 37.1\% & 44.8\% & 50.6\% & 53.2\% & 65.1\% & 68.5\% \\
\cmidrule(lr){1-1} \cmidrule(lr){2-4} \cmidrule(lr){5-7} \cmidrule(lr){8-10}
TA$^*$ & 25.6\% & 28.8\% & 31.4\% & 46.1\% & 48.7\% & 52.5\% & 64.7\% & 75.6\% & 79.4\% \\
\cmidrule(lr){1-1} \cmidrule(lr){2-4} \cmidrule(lr){5-7} \cmidrule(lr){8-10}
C$^*$ & 27.5\% & 28.8\% & 33.3\% & 47.4\% & 51.9\% & 55.1\% & 66.3\% & 75.6\% & 80.1\% \\
\cmidrule(lr){1-1} \cmidrule(lr){2-4} \cmidrule(lr){5-7} \cmidrule(lr){8-10}
C+TA$^*$  & 32.0\% & 35.2\% & 39.1\% & 54.4\% & 58.3\% & 62.1\% & 69.8\% & 79.4\% & 82.0\% \\
\bottomrule
\bottomrule
\multicolumn{13}{l}{$^*$: Base: Baseline decoding using top-$k$, TA: Temperature Adaptive Only,} \\
\multicolumn{13}{l}{~~~C: Contrastive Only, and C+TA: Contrastive with Temperature Adaptive.} \\
\multicolumn{13}{l}{$^+$: Pass@i means functional correctness in ``i'' runs.} \\
\end{tabular}
\end{table}

\subsection{Comparison with Base Decoding Techniques}

To show the overall effectiveness of our decoding approach, we evaluate it on three LLMs: (i) CodeLlama 7B (pre-trained), QwenCoder-2.5 14B (pre-trained), and CodeV (fine-tuned using RTL codes based on CodeQwen-2.5 7B). Our analysis focuses on two main metrics for RTL generation quality:

\begin{itemize}[leftmargin=*]
    \item \textit{Functional Correctness}: If the generated RTL behaves according to its specification, tested via logic simulation.
    \item \textit{Synthesizability}: If the RTL code can be successfully parsed, elaborated, and synthesized by a Verilog toolchain (Xilinx Vivado in our case), showing syntactic soundness.
\end{itemize}

In all experiments, we use a contrastive candidate set size of $k$ = 5, a contrastive penalty coefficient of $\lambda$ = 0.5, and a base decoding temperature of 0.7. As summarized in Table~\ref{tab:method_comparison_func} and Table~\ref{tab:method_comparison_synth} , our method outperforms standard decoding strategies, e.g., top-$k$ sampling, showing that our proposed decoding framework significantly reduces functionality and synthesizability issues, consistently leading to higher success rate in both perspectives.

\subsection{Computational Efficiency of Contrastive Decoding}

We evaluated the computational efficiency of our contrastive decoding strategy on the QwenCoder-2.5-14B model. To ensure a fair comparison of computational efficiency across decoding strategies, we enabled Key-Value (KV) caching during inference. KV caching significantly accelerates autoregressive decoding by storing the intermediate key and value tensors computed at each step of generation. This allows the model to avoid recomputing the entire attention mechanism over the full sequence at each step and instead perform attention only over the new token, thereby reducing inference time from quadratic to linear complexity with respect to sequence length. As shown in Table~\ref{tab:efficiency_qwen}, the average decoding time per token increased marginally from 0.1383 to 0.1413 seconds when using contrastive re-ranking. Peak GPU memory usage remained nearly unchanged, with only a 0.5\% increase.

In our proposed contrastive decoding strategy, at each decoding step, we sample a small set of candidate tokens (e.g., top-3) and compute their modified scores by combining the model's predicted log-probabilities with a contrastive penalty based on semantic similarity. This re-ranking is efficient for two reasons, (i) it operates on a small subset of tokens, not the full vocabulary, limiting the cost of additional computation, (ii) the similarity computations are lightweight vector operations (i.e., cosine similarity) between the precomputed token embeddings and the current context representation, avoiding any need for extra forward passes through the model. Because the token embeddings are already available at decoding time, the re-ranking step integrates seamlessly without disrupting the KV cache or triggering redundant computations.

\begin{table}[H]
\centering
\footnotesize
\setlength{\tabcolsep}{7pt}
\caption{Decoding Efficiency Comparison on QwenCoder-2.5-14B.}
\label{tab:efficiency_qwen}
\begin{tabular}{lcc}
\toprule
\textbf{Method} & \textbf{Avg. Time per Token (s)} & \textbf{Peak Memory (MB)} \\
\cmidrule(lr){1-1} \cmidrule(lr){2-2} \cmidrule(lr){3-3} 
Baseline        & 0.1383 & 15012.32 \\
\cmidrule(lr){1-1} \cmidrule(lr){2-2} \cmidrule(lr){3-3} 
Contrastive     & 0.1413 & 15089.43 \\
\bottomrule
\end{tabular}
\end{table}

\section{Conclusion}

In this paper, we investigated a critical yet underexamined dimension of RTL code generation with LLMs: the role of decoding strategies. While prior efforts have primarily focused on prompt engineering and fine-tuning, we demonstrate that decoding-time adaptations offer a lightweight and effective alternative for improving generation quality, without requiring additional data, model retraining, or architectural modifications. Through empirical analysis of entropy patterns and contextual token structures, we identify key challenges in RTL generation, including localized uncertainty and rigid syntactic dependencies. In response, we propose a domain-specific decoding framework that operates entirely at inference time, combining two complementary techniques: (1) a contrastive self-consistency mechanism that re-ranks top-$k$ candidates based on embedding-level diversity, encouraging confident yet diverse token selection; and (2) a syntax-aware temperature adaptation strategy that modulates sampling temperature based on the syntactic class of each token, effectively balancing determinism in structural regions with exploration in semantically rich contexts. Extensive experiments across three competitive LLMs, i.e., CodeLlama, QwenCoder, and CodeV, show that our method significantly improves both functional correctness and synthesizability, while also reducing hallucinations and repetition in the generated RTL code.

\bibliographystyle{IEEEtran}
\bibliography{references}

\begin{thebibliography}{10}
\providecommand{\url}[1]{#1}
\csname url@samestyle\endcsname
\providecommand{\newblock}{\relax}
\providecommand{\bibinfo}[2]{#2}
\providecommand{\BIBentrySTDinterwordspacing}{\spaceskip=0pt\relax}
\providecommand{\BIBentryALTinterwordstretchfactor}{4}
\providecommand{\BIBentryALTinterwordspacing}{\spaceskip=\fontdimen2\font plus
\BIBentryALTinterwordstretchfactor\fontdimen3\font minus \fontdimen4\font\relax}
\providecommand{\BIBforeignlanguage}[2]{{%
\expandafter\ifx\csname l@#1\endcsname\relax
\typeout{** WARNING: IEEEtran.bst: No hyphenation pattern has been}%
\typeout{** loaded for the language `#1'. Using the pattern for}%
\typeout{** the default language instead.}%
\else
\language=\csname l@#1\endcsname
\fi
#2}}
\providecommand{\BIBdecl}{\relax}
\BIBdecl

\bibitem{decortl_github}
{DecoRTL Repository - A Run-time Decoding Framework for RTL Code Generation with LLMs}, \url{https://github.com/mhakyash/DecoRTL}, 2025.

\bibitem{huang2021machine}
G.~Huang, J.~Hu, Y.~He, J.~Liu, M.~Ma, Z.~Shen, J.~Wu, Y.~Xu, H.~Zhang, K.~Zhong \emph{et~al.}, ``Machine learning for electronic design automation: A survey,'' \emph{ACM Transactions on Design Automation of Electronic Systems (TODAES)}, vol.~26, no.~5, pp. 1--46, 2021.

\bibitem{wu2024chateda}
H.~Wu, Z.~He, X.~Zhang, X.~Yao, S.~Zheng, H.~Zheng, and B.~Yu, ``Chateda: A large language model powered autonomous agent for eda,'' \emph{IEEE Transactions on Computer-Aided Design of Integrated Circuits and Systems}, 2024.

\bibitem{chang2023chipgpt}
K.~Chang, Y.~Wang, H.~Ren, M.~Wang, S.~Liang, Y.~Han, H.~Li, and X.~Li, ``Chipgpt: How far are we from natural language hardware design,'' \emph{arXiv preprint arXiv:2305.14019}, 2023.

\bibitem{akyash2024evolutionary}
M.~Akyash and H.~M~Kamali, ``Evolutionary large language models for hardware security: A comparative survey,'' in \emph{Proceedings of the great lakes symposium on VLSI 2024}, 2024, pp. 496--501.

\bibitem{akyash2025rtl++}
M.~Akyash, K.~Azar, and H.~Kamali, ``Rtl++: Graph-enhanced llm for rtl code generation,'' \emph{arXiv preprint arXiv:2505.13479}, 2025.

\bibitem{cui2024origen}
F.~Cui, C.~Yin, K.~Zhou, Y.~Xiao, G.~Sun, Q.~Xu, Q.~Guo, D.~Song, D.~Lin, X.~Zhang \emph{et~al.}, ``Origen: Enhancing rtl code generation with code-to-code augmentation and self-reflection,'' \emph{arXiv preprint arXiv:2407.16237}, 2024.

\bibitem{liu2024rtlcoder}
S.~Liu, W.~Fang, Y.~Lu, J.~Wang, Q.~Zhang, H.~Zhang, and Z.~Xie, ``Rtlcoder: Fully open-source and efficient llm-assisted rtl code generation technique,'' \emph{IEEE Transactions on Computer-Aided Design of Integrated Circuits and Systems}, 2024.

\bibitem{zhu2024hot}
Y.~Zhu, J.~Li, G.~Li, Y.~Zhao, Z.~Jin, and H.~Mei, ``Hot or cold? adaptive temperature sampling for code generation with large language models,'' in \emph{Proceedings of the AAAI Conference on Artificial Intelligence}, vol.~38, no.~1, 2024, pp. 437--445.

\bibitem{akyash2025simeval}
\BIBentryALTinterwordspacing
M.~Akyash and H.~Mardani~Kamali, ``Simeval: Investigating the similarity obstacle in llm-based hardware code generation,'' in \emph{Proceedings of the 30th Asia and South Pacific Design Automation Conference}, ser. ASPDAC '25.\hskip 1em plus 0.5em minus 0.4em\relax New York, NY, USA: Association for Computing Machinery, 2025, p. 1002–1007. [Online]. Available: \url{https://doi.org/10.1145/3658617.3697624}
\BIBentrySTDinterwordspacing

\bibitem{liu2023verilogeval}
M.~Liu, N.~Pinckney, B.~Khailany, and H.~Ren, ``Verilogeval: Evaluating large language models for verilog code generation,'' in \emph{2023 IEEE/ACM International Conference on Computer Aided Design (ICCAD)}.\hskip 1em plus 0.5em minus 0.4em\relax IEEE, 2023, pp. 1--8.

\bibitem{chen2021evaluating}
\BIBentryALTinterwordspacing
{M. Chen \emph{et al.}}, ``Evaluating large language models trained on code,'' 2021. [Online]. Available: \url{https://arxiv.org/abs/2107.03374}
\BIBentrySTDinterwordspacing

\bibitem{nijkamp2022codegen}
E.~Nijkamp, B.~Pang, H.~Hayashi, L.~Tu, H.~Wang, Y.~Zhou, S.~Savarese, and C.~Xiong, ``Codegen: An open large language model for code with multi-turn program synthesis,'' \emph{arXiv preprint arXiv:2203.13474}, 2022.

\bibitem{li2022competition}
\BIBentryALTinterwordspacing
{Y. Li \emph{et al.}}, ``Competition-level code generation with alphacode,'' \emph{Science}, vol. 378, no. 6624, p. 1092–1097, Dec. 2022. [Online]. Available: \url{http://dx.doi.org/10.1126/science.abq1158}
\BIBentrySTDinterwordspacing

\bibitem{Vijayakumar2018beam}
\BIBentryALTinterwordspacing
A.~Vijayakumar, M.~Cogswell, R.~Selvaraju, Q.~Sun, S.~Lee, D.~Crandall, and D.~Batra, ``Diverse beam search for improved description of complex scenes,'' \emph{Proceedings of the AAAI Conference on Artificial Intelligence}, vol.~32, no.~1, Apr. 2018. [Online]. Available: \url{https://ojs.aaai.org/index.php/AAAI/article/view/12340}
\BIBentrySTDinterwordspacing

\bibitem{fan2018hierarchical}
\BIBentryALTinterwordspacing
A.~Fan, M.~Lewis, and Y.~Dauphin, ``Hierarchical neural story generation,'' in \emph{Proceedings of the 56th Annual Meeting of the Association for Computational Linguistics (Volume 1: Long Papers)}, I.~Gurevych and Y.~Miyao, Eds.\hskip 1em plus 0.5em minus 0.4em\relax Melbourne, Australia: Association for Computational Linguistics, Jul. 2018, pp. 889--898. [Online]. Available: \url{https://aclanthology.org/P18-1082/}
\BIBentrySTDinterwordspacing

\bibitem{huang2024towards}
H.~Huang, Z.~Lin, Z.~Wang, X.~Chen, K.~Ding, and J.~Zhao, ``Towards llm-powered verilog rtl assistant: Self-verification and self-correction,'' \emph{arXiv preprint arXiv:2406.00115}, 2024.

\bibitem{yao2024rtlrewriter}
X.~Yao, Y.~Wang, X.~Li, Y.~Lian, R.~Chen, L.~Chen, M.~Yuan, H.~Xu, and B.~Yu, ``Rtlrewriter: Methodologies for large models aided rtl code optimization,'' \emph{arXiv preprint arXiv:2409.11414}, 2024.

\bibitem{akyash2024selfhwdebug}
M.~Akyash and H.~M. Kamali, ``Self-hwdebug: Automation of llm self-instructing for hardware security verification,'' in \emph{2024 IEEE Computer Society Annual Symposium on VLSI (ISVLSI)}, 2024, pp. 391--396.

\bibitem{mashnoor2025llmift}
N.~Mashnoor, M.~Akyash, H.~Kamali, and K.~Azar, ``Llm-ift: Llm-powered information flow tracking for secure hardware,'' in \emph{2025 IEEE 43rd VLSI Test Symposium (VTS)}, 2025, pp. 1--5.

\bibitem{zhang2024mg}
Y.~Zhang, Z.~Yu, Y.~Fu, C.~Wan, and Y.~C. Lin, ``Mg-verilog: Multi-grained dataset towards enhanced llm-assisted verilog generation,'' in \emph{2024 IEEE LLM Aided Design Workshop (LAD)}.\hskip 1em plus 0.5em minus 0.4em\relax IEEE, 2024, pp. 1--5.

\bibitem{liu2024craftrtl}
M.~Liu, Y.-D. Tsai, W.~Zhou, and H.~Ren, ``Craftrtl: High-quality synthetic data generation for verilog code models with correct-by-construction non-textual representations and targeted code repair,'' \emph{arXiv preprint arXiv:2409.12993}, 2024.

\bibitem{zhao2024codev}
Y.~Zhao, D.~Huang, C.~Li, P.~Jin, Z.~Nan, T.~Ma, L.~Qi, Y.~Pan, Z.~Zhang, R.~Zhang \emph{et~al.}, ``Codev: Empowering llms for verilog generation through multi-level summarization,'' \emph{arXiv preprint arXiv:2407.10424}, 2024.

\bibitem{thakur2024verigen}
S.~Thakur, B.~Ahmad, H.~Pearce, B.~Tan, B.~Dolan-Gavitt, R.~Karri, and S.~Garg, ``Verigen: A large language model for verilog code generation,'' \emph{ACM Transactions on Design Automation of Electronic Systems}, vol.~29, no.~3, pp. 1--31, 2024.

\bibitem{pei2024betterv}
Z.~Pei, H.-L. Zhen, M.~Yuan, Y.~Huang, and B.~Yu, ``Betterv: Controlled verilog generation with discriminative guidance,'' \emph{arXiv preprint arXiv:2402.03375}, 2024.

\bibitem{brown2020language}
T.~Brown, B.~Mann, N.~Ryder, M.~Subbiah, J.~D. Kaplan, P.~Dhariwal, A.~Neelakantan, P.~Shyam, G.~Sastry, A.~Askell \emph{et~al.}, ``Language models are few-shot learners,'' \emph{Advances in neural information processing systems}, vol.~33, pp. 1877--1901, 2020.

\bibitem{song2024good}
Y.~Song, G.~Wang, S.~Li, and B.~Y. Lin, ``The good, the bad, and the greedy: Evaluation of llms should not ignore non-determinism,'' \emph{arXiv preprint arXiv:2407.10457}, 2024.

\bibitem{renze2024effect}
M.~Renze, ``The effect of sampling temperature on problem solving in large language models,'' in \emph{Findings of the Association for Computational Linguistics: EMNLP 2024}, 2024, pp. 7346--7356.

\bibitem{holtzman2020curious}
\BIBentryALTinterwordspacing
A.~Holtzman, J.~Buys, L.~Du, M.~Forbes, and Y.~Choi, ``The curious case of neural text degeneration,'' 2020. [Online]. Available: \url{https://arxiv.org/abs/1904.09751}
\BIBentrySTDinterwordspacing

\bibitem{chuang2023dola}
Y.-S. Chuang, Y.~Xie, H.~Luo, Y.~Kim, J.~Glass, and P.~He, ``Dola: Decoding by contrasting layers improves factuality in large language models,'' \emph{arXiv preprint arXiv:2309.03883}, 2023.

\bibitem{su2022contrastive}
\BIBentryALTinterwordspacing
Y.~Su, T.~Lan, Y.~Wang, D.~Yogatama, L.~Kong, and N.~Collier, ``A contrastive framework for neural text generation,'' 2022. [Online]. Available: \url{https://arxiv.org/abs/2202.06417}
\BIBentrySTDinterwordspacing

\bibitem{li2023contrastive}
\BIBentryALTinterwordspacing
X.~L. Li, A.~Holtzman, D.~Fried, P.~Liang, J.~Eisner, T.~Hashimoto, L.~Zettlemoyer, and M.~Lewis, ``Contrastive decoding: Open-ended text generation as optimization,'' 2023. [Online]. Available: \url{https://arxiv.org/abs/2210.15097}
\BIBentrySTDinterwordspacing

\bibitem{o2023contrastive}
S.~O'Brien and M.~Lewis, ``Contrastive decoding improves reasoning in large language models,'' \emph{arXiv preprint arXiv:2309.09117}, 2023.

\bibitem{hui2024qwen2}
B.~Hui, J.~Yang, Z.~Cui, J.~Yang, D.~Liu, L.~Zhang, T.~Liu, J.~Zhang, B.~Yu, K.~Lu \emph{et~al.}, ``Qwen2. 5-coder technical report,'' \emph{arXiv preprint arXiv:2409.12186}, 2024.

\bibitem{roziere2023code}
B.~Roziere, J.~Gehring, F.~Gloeckle, S.~Sootla, I.~Gat, X.~E. Tan, Y.~Adi, J.~Liu, R.~Sauvestre, T.~Remez \emph{et~al.}, ``Code llama: Open foundation models for code,'' \emph{arXiv preprint arXiv:2308.12950}, 2023.

\end{thebibliography}

\end{document}